\documentclass[a4paper, notoc]{JHEP3}
\usepackage{amsfonts}
\usepackage{amssymb}
\usepackage{epsfig}
\usepackage{cite}

\newcommand{\be}{\begin{equation}}
\newcommand{\ee}{\end{equation}}
\newcommand{\bea}{\begin{eqnarray}}
\newcommand{\eea}{\end{eqnarray}}
\newcommand{\mbb}{\mathbb}
\newcommand{\ti}{\times}
\newcommand{\half}{\frac{1}{2}}
\newcommand{\mc}{\mathcal}

\newcommand{\vphi}{\varphi}

\title{Gaugino and Scalar Masses in the Landscape}
\author{Joseph P. Conlon, Fernando Quevedo\\ DAMTP, Centre for Mathematical
Sciences,\\
  Wilberforce Road, Cambridge, CB3 0WA, UK.\\

  E-mail:
  \email{j.p.conlon@damtp.cam.ac.uk} ,  \email{f.quevedo@damtp.cam.ac.uk}}
\abstract{ In this letter we demonstrate the genericity of suppressed gaugino masses
$ M_a \sim \frac{m_{3/2}}{\ln
(M_{Planck}/m_{3/2})} $ in the IIB string landscape, by showing that
this relation holds for D7-brane gauginos whenever the associated
modulus is stabilised by nonperturbative effects. Although
$m_{3/2}$ and $M_a$ take many different values across the landscape,
the above small mass hierarchy is maintained. We
  show that it is valid  
for models with an arbitrary number of moduli and applies to both the KKLT
and exponentially large volume  approaches to K\"ahler
moduli stabilisation. In the latter case we explicitly calculate
gaugino and moduli masses for compactifications on the two-modulus
Calabi-Yau $\mbb{P}^4_{[1,1,1,6,9]}$. In the large-volume
  scenario we also show that 
soft scalar masses are approximately universal with $m_i^2 \sim m_{3/2}^2(1 +
  \epsilon_i)$, with the non-universality parametrised by $\epsilon_i \sim
  \frac{1}{\ln (M_P/m_{3/2})^2} \sim \frac{1}{1000}$.
 We briefly discuss possible
  phenomenological implications of our results.}

\preprint{DAMTP-2006-37 \\ hep-th/0605141}

\begin{document}

\section{Introduction}

One of the most important problems in string theory is to make contact
with low-energy phenomenology. In this regard, the recent progress
made in moduli stabilisation represents a substantial advance
\cite{hepth0105097, hepth0301240}. Once
the moduli potential has been calculated, it becomes possible to study
in a top-down fashion supersymmetry breaking and, with suitable
assumptions about the loci of matter fields, the structure of soft
terms such as gaugino or scalar masses.

On the other hand, the same advances in moduli stabilisation have led
to a concern that the large discrete degeneracies present destroy any
possibility of low-energy predictivity. There seem to be many consistent
discrete choices for the fluxes that must be present (a standard
estimate is $10^{500}$). The worry is that the huge number of possible
choices will wash out any low-energy signals of high-energy physics.
This set of ideas and problems is encapsulated in the word `landscape'.

On a technical level, we use `landscape' to refer to IIB 
compactifications with moduli stabilised by fluxes and non-perturbative superpotential effects.
We shall work in this framework and use the formulae appropriate to it.
The dilaton and complex structure moduli are stabilised by 3-form
fluxes, while the K\"ahler moduli are stabilised by nonperturbative
superpotential effects. 
This is true in both the KKLT case
\cite{hepth0301240} and the exponentially large volume scenario developed
in \cite{hepth0502058}.
In principle, purely perturbative stabilisation may be
possible. However we shall
not consider this as no fully explicit examples exist even in a one
modulus case.

The first aim of this paper is to show that in the IIB landscape,
whenever a modulus $T_a$ is stabilised by nonperturbative effects there is a
small hierarchy between the masses of the gravitino $m_{3/2}$ and the
associated D7 gaugino $M_a$,
\be
\label{GauginoRelation}
M_a^2 \sim \frac{m_{3/2}^2}{\ln(m_{3/2})^2}.
\ee
We use natural units with $M_{Planck} = 2.4 \ti 10^{18} \hbox{GeV} \equiv 1$.
This small hierarchy was first identified in single-modulus
KKLT models \cite{hepth0411066} (with a two-modulus example also studied in
\cite{hepph0511162}). For the KKLT case this suppression of $M_a$ leads to mixed
modulus-anomaly mediation and
the phenomenology of this scenario
has been analysed in \cite{hepth0503216, hepph0504036, hepph0504037,
  hepph0507110, hepph0508029, hepph0509039, hepth0509158,
  hepth0511320, hepth0603047, hepph0604192, hepph0604253}.\footnote{In these
  references this hierarchy is present  for all soft supersymmetry
  breaking terms. We can establish it in generality only for
  gaugino masses. As we discuss in section \ref{secSM}, for the exponentially
  large volume scenario scalar masses are generically $\mc{O}(m_{3/2})$.} 
Here
we show this relation to be a truly
generic feature of the landscape, by showing it to hold for arbitrary
multi-modulus models and to be independent of
the precise details of the scenario used to stabilise the
moduli. The fact that (\ref{GauginoRelation}) relates
 $M_a$ to $m_{3/2}$ and $M_P$ emphasises that
it is a relation only among observable quantities and is independent of
the values of the fluxes that determine the structure of the landscape.

In section \ref{secSM} we also demonstrate soft scalar mass universality for
the exponentially large volume scenario and estimate the fractional
non-universality to be $1/\ln (m_{3/2})^2 \sim 1/1000$.
This is an interesting result as flavour non-universality is usually a
serious problem for gravity-mediated models. We briefly discuss the
phenomenology but defer a detailed discussion to \cite{ToCome}.

\section{Gaugino Masses}

For IIB flux compactifications
the appropriate K\"ahler potential and superpotential are given by
\bea
\label{KahlerPot}
\mc{K} & = & - 2 \ln \left( \mc{V} + \frac{\hat{\xi}}{2} \right),
\nonumber \\
\label{WDef}
W & = & W_0 + \sum A_i e^{-a_i T_i}.
\eea
Here $\mc{V} = \frac{1}{6}k_{ijk}t^i t^j t^k$ is the Calabi-Yau
volume, $\hat{\xi} = - \frac{\chi(M) \zeta(3)}{2(2 \pi)^3
g_s^{3/2}}$
and $T_i = \tau_i + i b_i$ are the K\"ahler moduli, corresponding to
the volume $\tau_i = \frac{\partial \mc{V}}{\partial t^i}$ of a
4-cycle $\Sigma_i$, complexified by the axion
$b_i = \int_{\Sigma_i} C_4$. $W_0 = \left\langle \int G_3 \wedge \Omega \right\rangle$ is the flux-induced superpotential
that is constant after integrating out dilaton and complex structure moduli.
In general there does not exist an explicit expression for $\mc{V}$ in
terms of the $T_i$.
We have included the perturbative K\"ahler corrections of
\cite{hepth0204254}. These are crucial in the exponentially large volume 
scenario but are not important for KKLT. We use single exponents in
the superpotential and do not consider
racetrack scenarios.

For a D7-brane wrapped on a cycle $\Sigma_i$, the gauge
kinetic function is given by
\be
f_i = \frac{T_i}{2\pi}.
\ee
Given the minimum of the moduli potential, the gaugino masses can be
computed through the general expression
\be
M_a = \half \frac{1}{\hbox{Re } f_a} \sum_\alpha F^\alpha \partial_\alpha f_a,
\ee
where $a$ runs over gauge group factors and  $\alpha$ over the moduli fields.
The F-term $F^\alpha$ is defined by
\bea
F^\alpha & = & e^{\mc{K}/2} \sum_{\bar{\beta}} \mc{K}^{\alpha \bar{\beta}} D_{\bar{\beta}}
\bar{W} \nonumber \\
& = & e^{\mc{K}/2} \sum_{\bar{\beta}} \mc{K}^{\alpha \bar{\beta}} \left( \partial_{\bar{\beta}}
\bar{W} + (\partial_{\bar{\beta}} \mc{K}) \bar{W} \right),
\eea
where we have expanded the covariant derivative
$D_{\bar{\beta}}
\bar{W} = \partial_{\bar{\beta}} \bar{W} + (\partial_{\bar{\beta}} \mc{K})
\bar{W}$.

Thus for a brane wrapping cycle $k$ we have
\be
\label{GauginoMass}
M_k = \half \frac{F^k}{\tau_k}.
\ee
It is a property of the K\"ahler potential $\mc{K} = - 2 \ln (\mc{V} +
\frac{\hat{\xi}}{2})$
that
\be
\label{Relation}
\sum_{\bar{j}} \mc{K}^{k \bar{j}} \partial_{\bar{j}} \mc{K} = -2
\tau_k \left(1 + \frac{\hat{\xi}}{4\mc{V}}\right) \equiv -2 \hat{\tau}_k.
\ee
We therefore obtain
\be
F^k = e^{\mc{K}/2} \left( \sum_{\bar{j}} \mc{K}^{k \bar{j}} \partial_{\bar{j}}
\bar{W} - (2 \hat{\tau}_k) \bar{W} \right).
\ee
From the superpotential (\ref{WDef}), we see that $\partial_{\bar{j}} \bar{W} = -a_j \bar{A_j} e^{-a_j \bar{T}_j}$,
and so
\be
\label{FTerm}
F^k = e^{\mc{K}/2} \left( \sum_{\bar{j}} - \mc{K}^{k \bar{j}} a_j \bar{A_j}
e^{-a_j \bar{T}_j} - 2 \hat{\tau}_k \bar{W} \right).
\ee
We now show that if the modulus $T_k$ is stabilised by
nonperturbative effects, the two terms in equation (\ref{FTerm})
cancel to leading order.
To see this, we start with the F-term supergravity potential,
\be
\label{Ftermpott}
V_F = e^{\mc{K}} \mc{K}^{i \bar{j}} \partial_i W \partial_{\bar{j}} \bar{W}
+ e^{\mc{K}} \mc{K}^{i \bar{j}} \left( (\partial_i K) W
\partial_{\bar{j}} \bar{W} + (\partial_{\bar{j}} \mc{K}) \bar{W}
\partial_i W \right) + e^{\mc{K}} (\mc{K}^{i \bar{j}} \mc{K}_i
\mc{K}_{\bar{j}} - 3) \vert W \vert^2.
\ee
Using (\ref{WDef}) and (\ref{Relation}), this becomes
\be
\label{ExpandedPotential}
V = \sum_{i \bar{j}} \frac{\mc{K}^{i \bar{j}} (a_i A_i) (a_j \bar{A}_j)
e^{-a_i T_i - a_j \bar{T}_j}}{\mc{V}^2} + \sum_j \frac{ 2 \hat{\tau}_j \left( a_j
\bar{A}_j e^{-a_j \bar{T}_j} W + a_j A_j \bar{W} e^{-a_j T_j}
\right)}{\mc{V}^2}
+ \frac{3 \hat{\xi} \vert W \vert^2}{4 \mc{V}^3}.
\ee
The perturbative K\"ahler corrections of (\ref{KahlerPot}) break no-scale,
giving the third term of (\ref{ExpandedPotential}).
We have only displayed the leading large-volume behaviour of these
corrections. This is reasonable as in KKLT such corrections are
not important, while in the exponentially large volume 
scenario $\mc{V} \gg 1$ and the higher volume-suppressed terms are
negligible.
We shall also assume throughout that $m_{3/2} \ll M_P$. This is
motivated by phenomenological applications and is any case necessary
to make sense of a small hierarchy governed by $\ln (M_P/m_{3/2})$.

We assume the modulus $T_k$ is stabilised by effects non-perturbative
in $T_k$
and locate the stationary locus by extremising
with respect to its real and imaginary parts. We first perform the
calculation keeping the dominant terms to demonstrate the cancellation
in (\ref{FTerm}). We subsequently show that the subleading terms are
indeed subleading and estimate their magnitude.

\subsection{Leading  Terms}

We first
solve for the axionic component,
$\partial V / \partial b_k = 0$. The axion only appears in the
superpotential and we have
\bea
\label{LongEquation}
\frac{\partial V}{\partial b_k} & = & \frac{i}{\mc{V}^2} \Bigg[
  \sum_{i \bar{j}} \mc{K}^{i \bar{j}} (a_i A_i)(a_j \bar{A}_j) \left[
    -a_i \delta_{i k} + a_j \delta_{j k} \right] e^{-(a_i T_i + a_j
    \bar{T}_j)} + \nonumber \\
& &
+ \sum_j 2 a_j \hat{\tau}_j \left( \bar{A}_j W a_j \delta_{jk} e^{-a_j
  \bar{T}_j} - A_j \bar{W} a_j \delta_{jk} e^{-a_j T_j} \right) \Bigg]
+ \\
& &  \sum_j \frac{ 2 \hat{\tau}_j \left( a_j
\bar{A}_j e^{-a_j \bar{T}_j} (\partial_{b_k} W) + a_j A_j (\partial_{b_k}
  \bar{W}) e^{-a_j T_j}
\right)}{\mc{V}^2} + \frac{3 \hat{\xi} ((\partial_{b_k} W) \bar{W} + W
  (\partial_{b_k} \bar{W}))}{4 \mc{V}^3}  \nonumber \\
& = & \frac{i}{\mc{V}^2} \Bigg[ - \sum_{j} \mc{K}^{k \bar{j}} (a_k^2
  A_k) (a_j \bar{A}_j) e^{-(a_k T_k + a_j \bar{T}_j)} + \sum_i
  \mc{K}^{i \bar{k}} (a_i A_i) (a_k^2 \bar{A}_k) e^{-(a_i T_i + a_k
    \bar{T}_k )} + \nonumber \\
& & + 2 a_k^2 \hat{\tau}_k \left( \bar{A}_k W e^{-a_k
    \bar{T}_k} - A_k \bar{W} e^{-a_k T_k} \right) \Bigg].
\label{ShorterEquation}
\eea
In going from (\ref{LongEquation}) to (\ref{ShorterEquation}) we have
dropped the third line of (\ref{LongEquation}) as subleading. We will
estimate the magnitude of these subleading terms in 
section \ref{SubTerms}.

We now change the dummy index in (\ref{ShorterEquation}) from $j$ to $i$, and use $\mc{K}^{k \bar{i}}
= \mc{K}^{i \bar{k}}$ together with $\frac{\partial V}{\partial b_k} =
0$ to obtain
\bea
\label{AxionEquation}
2 \hat{\tau}_k (\bar{A}_k W e^{-a_k \bar{T}_k} - A_k \bar{W} e^{-a_k
  T_k}) & = &
\sum_i \mc{K}^{k \bar{i}} \left( (a_i \bar{A}_i) A_k  e^{-(a_k T_k +
  a_i \bar{T}_i)} - (a_i A_i) \bar{A}_k e^{-(a_k \bar{T}_k +a_i T_i)}
\right)  \nonumber \\
& & + (\hbox{subleading terms}).
\eea
As the axion does not appear (at least in perturbation theory) in the
K\"ahler potential, its stabilisation is always entirely due to
  nonperturbative superpotential effects.

We next consider the stabilisation of $\tau_k = \hbox{Re}(T_k)$.
As stated above, our main assumption is that $T_k$ is stabilised by
superpotential effects nonperturbative in $T_k$. Another way of stating this
is to say that, when computing $\frac{\partial V}{\partial \tau_k}$,
the dominant contribution must arise from the superpotential term $A_k
e^{-a_k T_k}$: if this were not the case, our assumption about how
$T_k$ is stabilised is invalid.
In evaluating $\frac{\partial V}{\partial \tau_k}$,
we therefore focus on such terms as dominant and neglect terms
arising from e.g. $\frac{\partial}{\partial \tau_k} \left( \frac{\mc{K}^{i
  \bar{j}}}{\mc{V}^2}\right)$ as subdominant. We show in
section \ref{SubTerms}
 that the magnitude of the subdominant terms is suppressed by factors of
$\ln \left( m_{3/2} \right)$.

If we only consider superpotential terms, the calculation of
$\frac{\partial V}{\partial \tau_k}$
exactly parallels that of $\frac{\partial V}{\partial b_k}$ above. The only
differences lie in the signs, as
$$
\frac{\partial T_k}{\partial \tau_k} = \frac{\partial \bar{T}_k}{\partial \tau_k}
= 1, \hbox{ whereas } \frac{\partial T_k}{\partial b_k} =
-\frac{\partial \bar{T}_k}{\partial b_k} = i.
$$
In a similar fashion to (\ref{ShorterEquation}) we therefore obtain
\bea
\label{Tderivative}
\frac{\partial V}{\partial \tau_k} & = & \frac{a_k^2}{\mc{V}^2} \Big[
  \sum_i \mc{K}^{k \bar{i}} \Big( - (a_i \bar{A}_i) A_k e^{-(a_k T_k
    +a_i \bar{T}_i)} -  (a_i A_i) \bar{A}_k e^{-(a_k \bar{T}_k +a_i
    T_i)} \Big) \nonumber \\
& &
- 2 \hat{\tau}_k \left( \bar{A}_k W e^{-a_k \bar{T}_k} +
  A_k \bar{W} e^{-a_k T_k} \right) \Big] + (\hbox{subleading terms}).
\eea
Setting $\frac{\partial V}{\partial \tau_k}=0$ then implies
\bea
\label{TauEquation}
2 \hat{\tau}_k \left( \bar{A}_k W e^{-a_k \bar{T}_k} + A_k \bar{W} e^{-a_k
  {T}_k} \right) & = & - \sum_i \mc{K}^{k \bar{i}} \left( a_i \bar{A}_i A_k
e^{-(a_k T_k + a_i \bar{T}_i)} + a_i A_i \bar{A}_k e^{-(a_k \bar{T}_k
  + a_i T_i)} \right) \nonumber  \\
& & + (\hbox{subleading terms}).
\eea
We now sum (\ref{AxionEquation}) and (\ref{TauEquation}) to obtain
\bea
4 \hat{\tau}_k \bar{A}_k W e^{-a_k \bar{T}_k} & = & -2 \sum_i \mc{K}^{k
  \bar{i}} a_i A_i \bar{A}_k e^{-(a_k \bar{T}_k + a_i T_i)} \nonumber \\
\Rightarrow -2 \hat{\tau}_k W & = & \sum_i \mc{K}^{k \bar{i}} a_i A_i
  e^{-a_i T_i} \nonumber \\
\label{LastEquation}
\Rightarrow - 2 \hat{\tau}_k \bar{W} & = & \sum_i \mc{K}^{\bar{k} i} a_i \bar{A}_i
  e^{-a_i \bar{T}_i}.
\eea
Comparison with equations (\ref{GauginoMass}) and (\ref{FTerm}) shows
  that there exists a leading-order cancellation in the computation of
  the gaugino mass. This cancellation has followed purely from the
  assumption that the modulus $\tau_k$ was stabilised by non-perturbative
  effects: we have only required $\frac{\partial V}{\partial \tau_k} =
  0$ and not $D_{T_k} W = 0$. 

In deriving equations (\ref{ShorterEquation}) and (\ref{Tderivative}) we dropped
  subleading terms suppressed by $\ln \left( \frac{M_P}{m_{3/2}}
  \right)$. We then expect the cancellation from (\ref{LastEquation})
  and (\ref{FTerm}) to
  fail at this order, giving
\be
F^k \sim - 2 \frac{\hat{\tau}_k e^{\mc{K}/2} \bar{W}}{\ln(m_{3/2})}.
\ee
Equation (\ref{GauginoMass}) then gives
\be
M_k \sim \frac{e^{\mc{K}/2}\bar{W}}{\ln(m_{3/2})} = \frac{m_{3/2}}{\ln (m_{3/2})},
\ee
the relation we sought.

The above argument is general and model-independent. We have used the
K\"ahler potential appropriate to an arbitrary compactification,
making no assumptions about the number of moduli. Furthermore, as the
result comes from directly extremising the scalar potential it is
independent of whether the moduli stabilisation is approximately
supersymmetric or not.
Indeed, the argument above has
  not depended on finding a global minimum of the potential, or even on
  extremising the potential with respect to any of the moduli except $T_k$.
This result shows that the small hierarchy of (\ref{GauginoRelation})
will exist in both
KKLT and exponentially large volume approaches to moduli
  stabilisation. In the latter case this is possibly
  surprising\footnote{and was to the authors.}, as
  the minimum is in no sense approximately susy: each contribution to
  the sum in (\ref{FTerm}) individually gives $M_a \sim m_{3/2}$: it
  is only when summed the mass suppression is obtained.

We note, as an aside, that if \emph{all} moduli are
  stabilised by non-perturbative effects then by contracting
  (\ref{LastEquation}) with $\mc{K}_{j \bar{k}}$ we obtain
\be
-2 \sum_k \mc{K}_{j \bar{k}} \hat{\tau}_k \bar{W} = a_j \bar{A}_j
  e^{-a_j \bar{T}_j}.
\ee
Now, $\mc{K}_{j}$ is homogeneous of degree -1 in $\tau_k$, so
  recalling that $\frac{\partial}{\partial \tau_k} = 2
  \frac{\partial}{\partial T_k}$,
$$
\sum_k -2 \mc{K}_{j \bar{k}} \tau_k = \sum_k - \frac{\partial
  \mc{K}_j}{\partial \tau_k} \tau_k = \mc{K}_j,
$$
and therefore to leading order
\be
\label{SusyStab}
\partial_j W + (\partial_j \mc{K}) \bar{W} = 0.
\ee
Consequently if \emph{all} moduli are stabilised by nonperturbative
effects then the stabilisation is always approximately supersymmetric.

\subsection{Subleading terms}
\label{SubTerms}

We now want to show that the terms neglected in computing
$\frac{\partial V}{\partial \tau_k}$ are all suppressed, under the
assumption that the modulus is solely stabilised by nonperturbative
effects. For concreteness we focus on the term in the potential
\be
\label{PotentialFirstTerm}
\sum_{i, \bar{j}} \left( \frac{\mc{K}^{i \bar{j}}}{\mc{V}^2} \right) (a_i A_i)(a_j \bar{A}_j)
  e^{-(a_i T_i + a_j \bar{T}_j)}.
\ee
The argument used for this term will also apply to the other terms of (\ref{ExpandedPotential}).
$ \frac{\mc{K}^{i \bar{j}}}{\mc{V}^2}$ is homogeneous in the $\tau_k$
of degree $-1$. To see this, we note that as by dimensional analysis
$\mc{V}$ is homogeneous in the $\tau_i$ of degree 3/2, $\mc{K}_{i
  \bar{j}} = \frac{\partial}{\partial T_i} \frac{\partial}{\partial
  \bar{T}_j} ( -2 \ln (\mc{V}))$
  is homogeneous in the $\tau_i$ of degree -2, and so $\mc{K}^{i
    \bar{j}}$ is homogeneous in the $\tau_i$ of degree 2. Therefore,
  summing over $k$,
\be
\sum_k \tau_k \frac{\partial}{\partial \tau_k} \left( \frac{\mc{K}^{i
    \bar{j}}}{\mc{V}^2} \right) = - \frac{\mc{K}^{i
    \bar{j}}}{\mc{V}^2},
\ee
and so
\be
\frac{\partial}{\partial \tau_k} \left( \frac{\mc{K}^{i
    \bar{j}}}{\mc{V}^2}\right) \lesssim \frac{\mc{K}^{i
    \bar{j}}}{\tau_k \mc{V}^2}.
\ee
Cosequently, differentiating (\ref{PotentialFirstTerm}) w.r.t $\tau_k$ gives
$$
\mc{O}\left( \frac{1}{\tau_k} \sum_{i,j} \frac{\left(\mc{K}^{i \bar{j}} (a_i A_i) (a_j
  \bar{A}_j) e^{-(a_i T_i + a_j \bar{T}_j)} \right)}{\mc{V}^2} \right)
+ a_k \sum_j \frac{\left(\mc{K}^{k \bar{j}} (a_k A_k) (a_j \bar{A}_j)
  e^{-(a_k T_k + a_j \bar{T}_j)} + c.c\right)}{\mc{V}^2}.
$$
The basic assumption we make is that the location of the minimum for
$\tau_k$ is dominantly determined by the effects nonperturbative in
$\tau_k$. Therefore in the first sum we should only include the terms which
depend nonperturbatively on $a_k T_k$. This gives
\be
\label{Rewrite}
\mc{O}\left( \frac{1}{\tau_k} \sum_j \frac{\left(\mc{K}^{k \bar{j}} (a_k A_k) (a_j \bar{A}_j)
  e^{-(a_k T_k + a_j \bar{T}_j)} + c.c\right)}{\mc{V}^2} \right)
+ a_k \sum_j \frac{\left(\mc{K}^{k \bar{j}} (a_k A_k) (a_j \bar{A}_j)
  e^{-(a_k T_k + a_j \bar{T}_j)} + c.c\right)}{\mc{V}^2}.
\ee
We then see that the first
term of (\ref{Rewrite}) is suppressed compared to the second by a factor $a_k
\tau_k$.

We note that there can exist moduli $\tau_k$ not stabilised
by effects nonperturbative in $\tau_k$. This certainly holds for the
volume modulus in the exponentially large volume models of
\cite{hepth0502058}. Furthermore, one can argue that that in order to avoid
generating a potential for the QCD axion, the modulus $\tau_k$ associated with the QCD cycle should
be stabilised without using effects nonperturbative in $\tau_k$. 
The relationship between moduli stabilisation and the existence of a
QCD axion is discussed further in \cite{hepth0602233} (also see \cite{hepth0309170}).
Our argument above is restricted to the
case where the modulus $\tau_k$ is stabilised by effects
nonperturbative in $\tau_k$.

An identical analysis applies to the other two terms of
equation (\ref{ExpandedPotential}). As the K\"ahler dependent terms
are polynomials in $\tau_k$, derivatives of these with respect to
$\tau_k$ also give a suppression factor of $\tau_k$, while the derivatives
of superpotential exponents are enhanced by a factor $a_k$. The latter (which we
keep) are therefore larger by a factor $a_k \tau_k$ than the terms discarded.

In passing from (\ref{LongEquation}) to (\ref{ShorterEquation}) we
dropped the last line of (\ref{LongEquation}). This is self-consistent
so long as $\sum_j A_j e^{-a_j T_j}$ is suppressed compared to $W$. In
the exponentially large-volume scenario this is trivial as $e^{-a_k
  \tau_k} \sim \frac{1}{\mc{V}}$ while $W \sim \mc{O}(1)$. In KKLT
models, as 
$$
\partial_{T_i} \mc{K} = -2 \frac{\partial_{T_i} \mc{V}}{\mc{V}} \lesssim
\frac{2}{\tau_i},
$$
we can use (\ref{SusyStab}) to see that
\be
\bar{W} \gtrsim (a_k \tau_k) A_k e^{-a_k \tau_k},
\ee
and so the third line of (\ref{LongEquation}) is suppressed compared
to the second by a factor of (at least) $a_k \tau_k$. 

The above arguments imply that the terms dropped in our evaluation
of $\frac{\partial V}{\partial \tau_k}$ either
\begin{enumerate}
\item are suppressed by a factor of $a_k \tau_k \sim \ln(m_{3/2})$ or
\item have no non-perturbative dependence on $\tau_k$.
\end{enumerate}
Consequently the gaugino mass suppression found above will hold at
leading order in $\frac{1}{\ln (m_{3/2})}$.

\subsection{The uplift term}

In order to attain almost vanishing but positive vacuum energy, an
uplift term must also be
included. In KKLT this is in a sense responsible for the soft
masses, as the original AdS minimum is supersymmetric. In the
exponentially large volume case the AdS minimum is already
non-supersymmetric and the contribution of the uplift to soft terms
is less relevant. 

We take as
uplift 
\be V_{uplift} = \frac{\epsilon}{\mc{V}^\alpha}, \ee 
where the power $4/3 \le \alpha \le 2$ depends on the uplift mechanism
\cite{hepth0301240, bkq,ss}.
Including
this phenomenological uplift term, the full potential is \be
\label{UpliftedPotential} V_{full} = V_{SUGRA} + V_{uplift}. \ee At
the minimum $\langle V_{full} \rangle = 0$ and so we must have
$$
\langle V_{SUGRA} \rangle = - \langle V_{uplift} \rangle.
$$
$\mc{V}^{-\alpha}$ is homogeneous of degree $-3\alpha/2$ in the $\tau_i$ and so
\be
\sum_k \tau_k \frac{\partial}{\partial \tau_k} \mc{V}^{-\alpha} =
-\frac{3\alpha}{2}  \mc{V}^{-\alpha},
\ee
implying
\be
\frac{\partial}{\partial \tau_k} \mc{V}^{-\alpha} \lesssim - \frac{3\alpha}{2\tau_k}
\mc{V}^{-\alpha}.
\ee
Thus $\frac{\partial V_{uplift}}{\partial \tau_k}$ is
suppressed compared to $V_{uplift}$ by a factor of $\tau_k$.
In contrast, the derivatives of $V_{SUGRA}$ involve an enhancement by
$a_k$ due to the exponentials. As the two terms of
(\ref{UpliftedPotential}) are by definition equal at the minimum, we
see that
\be
\frac{\partial V_{SUGRA}}{\partial \tau_k} \gtrsim a_k \tau_k
\frac{\partial V_{uplift}}{\partial \tau_k}.
\ee
This fits in with
our previous analysis: the terms giving rise to the cancellation in (\ref{FTerm})
are the leading ones, with subleading terms suppressed by $a_k
\tau_k$. We do not control the subleading terms and they will
generically be non-vanishing, giving further contributions to the gaugino
masses at $\mc{O}\left(\frac{m_{3/2}}{\ln(m_{3/2})}\right)$.

As an aside, we note that for the exponentially large volume models
\be
\frac{\partial V_{uplift}}{\partial \tau_k} \sim \frac{1}{\mc{V}}
V_{uplift},
\ee
and so the presence of the uplift term does not significantly affect
the stabilisation of $\tau_k$. 

\section{Explicit Calculation for $\mbb{P}^4_{[1,1,1,6,9]}$}

The above has established that a suppression of gaugino masses
compared to the gravitino mass is generic in the landscape.
We now illustrate the above with explicit calculations for
compactifications on $\mbb{P}^4_{[1,1,1,6,9]}$ in the
exponentially large volume scenario.

\subsection{Gaugino Masses}

We calculate the gaugino masses explicitly for exponentially large volume
flux compactifications on $\mbb{P}^4_{[1,1,1,6,9]}$. We
briefly recall the relevant properties of the model - a fully detailed
analysis can be found in
\cite{hepth0502058, hepth0505076}. As the manifold has $h^{1,1}=2$, there are two moduli, $T_b$ and
$T_s$. $T_b$ controls the overall volume and $T_s$ is a blow-up mode.
At the minimum $\tau_b = \hbox{Re}(T_b) \gg \tau_s = \hbox{Re}(T_s)
\sim \ln(\tau_b)$.\footnote{The prefixes $b$ and $s$ indicate `big' and
  `small' moduli. It should be properly understood that this means that
  $\tau_b$ is an exponentially large modulus determining the overall
  volume, whereas $\langle \tau_s \rangle$ is smaller than $\langle\tau_b\rangle$
  but still larger than the string scale, and so the low-energy 4D
  effective theory is justified.}
The K\"ahler and superpotential are
given by\footnote{For simplicity we do not include a factor of
  $\frac{1}{9\sqrt{2}}$ in $\mc{V}$: this does not affect the results.}
\bea
\mc{K} & = & - 2 \ln \left( \mc{V} + \frac{\hat{\xi}}{2} \right) \equiv -
2 \ln \left( \tau_b^{3/2} - \tau_s^{3/2} + \frac{\hat{\xi}}{2}
\right), \\
W & = & W_0 + A_s e^{-a_s T_s} + A_b e^{-a_b T_b}.
\eea
The resulting K\"ahler metric is (dropping terms subleading in powers
of $\mc{V}$)
\be
\label{Metric}
\mc{K}_{i \bar{j}} = \left( \begin{array}{cc} \mc{K}_{b \bar{b}} &
  \mc{K}_{b \bar{s}} \\ \mc{K}_{s \bar{b}} & \mc{K}_{s \bar{s}}
\end{array} \right) = \left( \begin{array}{cc} \frac{3}{4 \mc{V}^{4/3}} &
  -\frac{9 \tau_s^{\half}}{8 \mc{V}^{5/3}} \\
-\frac{9 \tau_s^{\half}}{8 \mc{V}^{5/3}} & \frac{3}{8 \sqrt{\tau_s}
  \mc{V}} \end{array} \right),
\ee
with inverse metric
\be
\label{InverseMetric}
\mc{K}^{i \bar{j}} =  \left( \begin{array}{cc} \mc{K}^{b \bar{b}} &
  \mc{K}^{b \bar{s}} \\ \mc{K}^{s \bar{b}} & \mc{K}^{s \bar{s}}
\end{array} \right) = \left( \begin{array}{cc} \frac{4 \mc{V}^{4/3}}{3} &
  4 \tau_s \tau_b \\
4 \tau_s \tau_b & \frac{8 \sqrt{\tau_s} \mc{V}}{3} \end{array} \right).
\ee

In a limit $\mc{V} \equiv \tau_b^{3/2} - \tau_s^{3/2} \gg 1$ with
$\tau_s \sim \mc{O}(1)$, direct
evaluation of the scalar potential gives (dropping terms
subleading in $\mc{V}$),
\be
\label{BBCQPotential}
V = \frac{\lambda a_s^2 A_s^2 \sqrt{\tau_s} e^{-2 a_s \tau_s}}{\mc{V}}
- \frac{\mu a_s A_s \tau_s \vert W_0 \vert e^{-a_s \tau_s}}{\mc{V}^2} +
\frac{\nu \vert W_0 \vert^2}{\mc{V}^3}.
\ee
Explicitly, $\lambda = \frac{8}{3}$ and $\mu = 4$. The minus sign in
(\ref{BBCQPotential}) comes from minimising the potential with respect
to the axion $b_s$. As
$\tau_b \gg 1$ all terms nonperturbative in $\tau_b$ vanish. Noting
that
 $\frac{\partial}{\partial \tau_s}\left( \mc{V}^{-1} \right) \sim
\mc{O}\left(\frac{1}{\mc{V}^2}\right)$, we obtain
\be
\label{aaa}
\frac{\partial V}{\partial \tau_s} = \frac{\lambda a_s^2 A_s^2 \sqrt{\tau_s}
  e^{-2 a_s \tau_s}}{\mc{V}} \left( -2 a_s + \frac{1}{2 \tau_s}
\right) - \frac{\mu a_s A_s e^{-a_s \tau_s} \vert W_0 \vert}{\mc{V}^2}
\left( -a_s \tau_s +1 \right) + \mc{O}\left( \frac{1}{\mc{V}^2}
\right).
\ee
Imposing $\frac{\partial V}{\partial \tau_s} = 0$ and rearranging (\ref{aaa})
gives
\be
\label{SmallModSolution}
e^{-a_s \tau_s} = \left( \frac{\mu}{2 \lambda}\right)\frac{\vert W_0
  \vert}{\mc{V} a_s} \sqrt{\tau_s} \left(1 - \frac{3}{4 a_s \tau_s}
  \right) + \mc{O}\left(\frac{1}{(a_s \tau_s)^2} \right).
\ee
If we also solve $\frac{\partial V}{\partial \mc{V}} = 0$, we obtain \cite{hepth0502058}
$$
\mc{V} \sim \left\vert \frac{W_0}{A_s} \right\vert e^{a_s \tau_s} \qquad \textrm{ with }
\tau_s \sim \hat{\xi}^{\frac{2}{3}}.
$$
$\mc{V}$ is exponentially sensitive to $\hat{\xi}$ (which includes
$g_s$) and $a_4$ and so can
take on essentially any value. A TeV scale gravitino mass requires
$\mc{V} \sim 10^{14}$, which we assume.

There are gaugini associated with the moduli $\tau_b$ and
$\tau_s$. From (\ref{GauginoMass}) these have masses $\frac{F^b}{2 \tau_b}$ and $\frac{F^s}{2
  \tau_s}$ respectively. We calculate both masses: however we note
that the small cycle is the only cycle appropriate for Standard Model
matter, as a brane wrapped on the large cycle would have too small a
gauge coupling.
First,
\bea
F^b & = & e^{\mc{K}/2} \left(\mc{K}^{b \bar{b}} D_{\bar{b}} \bar{W} + \mc{K}^{b
  \bar{s}} D_{\bar{s}} \bar{W} \right) \\
& = & e^{\mc{K}/2} \left( -2 \tau_b \bar{W} + \mc{K}^{b \bar{b}}
\partial_{\bar{b}} \bar{W} + \mc{K}^{b \bar{s}} \partial_{\bar{s}}
\bar{W} \right),
\eea
where we have used $\mc{K}^{b \bar{b}} \partial_{\bar{b}} \mc{K} +
\mc{K}^{b \bar{s}}\partial_{\bar{s}} \mc{K} = -2 \tau_b$ from (\ref{Relation}).
However, as $\tau_b \sim \mc{V}^{2/3} \gg 1$, $\partial_{\bar{b}}
\bar{W} \sim \exp (-a_b \tau_b) \sim 0$. Furthermore,
$$
\mc{K}^{b \bar{s}} \partial_{\bar{s}} \bar{W} \sim (4 \tau_b \tau_s) \times
(a_s A_s \exp(-a_s \tau_s)) \sim \mc{V}^{-1/3},
$$
and so
\be
F^b = \frac{1}{\mc{V}} \left( -2 \tau_b W_0 +
\mc{O}\left(\mc{V}^{-1/3}\right)\right),
\ee
implying
\be
\vert M_b \vert = \left\vert \frac{F^b}{2 \tau_b} \right\vert = m_{3/2} + \mc{O}(\mc{V}^{-1/3}).
\ee
This is an identical relation to that of the fluxed MSSM \cite{hepph0408064}.

We now calculate $M_s$. Using $\mc{K}^{i \bar{j}} \mc{K}_{\bar{j}} =
- 2 \tau_i$ and $\mc{K}^{s \bar{b}} \partial_{\bar{b}} W \sim 0$, we get
\bea
F^s & = & e^{\mc{K}/2} \left( \mc{K}^{s \bar{s}} \partial_s \bar{W} - 2
\tau_s \bar{W} \right) \\
& = & e^{\mc{K}/2} \left( \mc{K}^{s \bar{s}} (-a_s A_s e^{-a_s T_s}) - 2
\tau_s \bar{W} \right).
\eea
From (\ref{InverseMetric}), $\mc{K}^{s \bar{s}} = \frac{8 \sqrt{\tau_s} \mc{V}}{3}$. Using
(\ref{SmallModSolution}) we then have
\be
F^s = \frac{2 \tau_s \bar{W}}{\mc{V}} \left( \left(1 - \frac{3}{4 a_s
  \tau_s} \right) - 1 \right).
\ee
We therefore obtain
\be
\label{SmallGauginoMass}
\vert M_s \vert = \frac{3 m_{3/2}}{4 a_s \tau_s}\left( 1+
\mc{O}\left(\frac{1}{a_4 \tau_4}\right)\right) = \frac{3 m_{3/2}}{4
  \ln (m_{3/2})} \left(1 + \mc{O}\left( \frac{1}{\ln(m_{3/2})}\right)\right),
\ee
with the expected small hierarchy. We therefore conclude that the gaugino mass associated to
 the exponentially large modulus $\tau_b$ is equal to the gravitino
 mass, whereas the gaugino mass associated to the small modulus
 $\tau_s$ is suppressed by $\ln (m_{3/2})$. While in the
 $\mbb{P}^4_{[1,1,1,6,9]}$ case there is only one
 small modulus, this result will extend
 to more realistic multi-modulus examples.
Notice that it is on the
 smaller cycles that the Standard Model should be accomodated since on
 the larger ones the gauge coupling is exponentially small and unrealistic.

\subsection{Moduli Masses}

We also here prove the existence of an enhancement by
$\ln(m_{3/2})$ in the mass of the small modulus $\tau_s$ compared to
$m_{3/2}$. This is similar behaviour as found in KKLT solutions \cite{hepth0503216}.\footnote{The
  large $vev$ modulus $\tau_b$ has a mass further suppressed by a
  factor of ${\cal V}^{1/2}$ and is therefore much lighter than
  $\tau_s$ \cite{hepth0505076}.} This extends the simple
volume scaling arguments of \cite{hepth0505076} which gave $m_s \sim \frac{M_P}{\mc{V}}$ and $m_b
\sim \frac{M_P}{\mc{V}^{3/2}}$. Focusing purely on the field
$\tau_s$, its Lagrangian is
\be
\int d^4 x \, \mc{K}_{s \bar{s}} \partial_\mu \tau_s \partial^\mu \tau_s + V(\tau_s).
\ee
As $\tau_s$ is the heavier of the two moduli, a lower bound on its
mass is given by
\be
m_s^2 \gtrsim \frac{1}{2 \mc{K}_{s\bar{s}}} \left\langle \frac{\partial^2
  V}{\partial \tau_s^2} \right\rangle.
\ee
Now, $\mc{K}_{s \bar{s}} = \frac{3}{8 \sqrt{\tau_s} \mc{V}}$. If we
evaluate $\frac{\partial^2 V}{\partial \tau_s^2}$, we obtain
\be
\frac{\partial^2 V}{\partial \tau_s^2} = \frac{4 \lambda a_s^4
  \sqrt{\tau_s} e^{-2 a_s \tau_s}}{\mc{V}} - \frac{\mu a_s^3 \tau_s
  e^{-a_s \tau_s} \vert W_0 \vert}{\mc{V}^2} + \left(\hbox{terms suppressed
  by }\frac{1}{a_s \tau_s}\right).
\ee
Substituting in our evaluation of $\langle e^{-a_s \tau_s} \rangle $
from (\ref{SmallModSolution}), we
obtain
\be
\frac{\partial^2 V}{\partial \tau_s^2} = \left( \frac{\mu^2}{2
  \lambda} \right) \frac{a_s^2 \tau_s^{3/2} \vert W_0
  \vert^2}{\mc{V}^3} \left(1 + \mc{O}\left(\frac{1}{a_s \tau_s}\right)\right)
\ee
Consequently
\bea
m_{\tau_s}^2 & \gtrsim & \left( \frac{4 \sqrt{\tau_s}\mc{V}}{3}\right)
\left( \frac{\mu^2}{2 \lambda} \right) \frac{a_s^2 \tau_s^{3/2} \vert
  W_0 \vert^2}{\mc{V}^3} \nonumber \\
& = & \left( \frac{2 \mu^2}{3 \lambda} \right) \frac{a_s^2 \tau_s^2
  \vert W_0 \vert^2}{\mc{V}^2}.
\eea
This gives
\be
\label{SmallModMass}
m_{\tau_s} \gtrsim 2 \ln \left( \frac{M_P}{m_{3/2}} \right) m_{3/2}
\left(1 + \mc{O}\left(\frac{1}{\ln(m_{3/2})}\right)\right).
\ee
While technically a lower bound, this is actually a very good estimate
of $m_{\tau_s}$, as the canonically normalised heavy modulus has only a very
small admixture of $\tau_b$.
This is confirmed by explicit numerical evaluation, which shows the
formulae  (\ref{SmallGauginoMass}) and (\ref{SmallModMass}) to be
accurate to within a couple of per cent.

\section{Scalar Masses}
\label{secSM}

The suppressed values for the
gaugino masses are a direct consequence of a cancellation in the 
calculation for the F-terms for the `small' moduli. Having suppresed
F-terms could naively lead to the conclusion that the
other soft terms must also be suppressed. This is not necessarily the
case. For example, the general expression for scalar masses  
depends explicitly on the form of the K\"ahler potential
for matter fields $\varphi$. Suppose we write
\be
\mc{K}(\vphi, \bar{\vphi}, T_m, \bar{T_n}) = \mc{K}_0 (T, \bar{T}) + 
\tilde{\mc{K}}(T_m,\bar{T}_n)\varphi \bar{\varphi} + \ldots,
\ee
where the index labelling different scalar fields
has been omitted. 
This
leads to the well known expression for scalar masses
\cite{luis}:\footnote{This form assumes the matter metric is diagonal: the
  results below are unaffected if we use the fully general expressions \cite{ToCome}.}
\be
\label{scalarmass}
m_{\vphi}^2 \ =\ m_{3/2}^2 + V_0 - F^m\bar{F}^{\bar n} \partial_m\partial_{\bar
  n}  (\ln \tilde{\cal K}).
\ee 
In order to get suppressed values for $m_{\vphi}^2$ there must be a
contribution cancelling the leading $m_{3/2}^2$
contribution (assuming a negligible vacuum energy
$V_0$). In the KKLT models, this is provided by the anti-D3 brane \cite{hepth0503216}.
However, this cancellation is not generic.

For the large-volume models the uplift term is subdominant in susy
breaking and so has no significant effect on (\ref{scalarmass}). As
the F-terms are suppressed by $\ln (m_{3/2})$ for all small moduli,
the only F-term that can cancel the gravitino mass contribution is
that associated to the large volume modulus. The dependence of
$\tilde{K}$ on $\mc{V}$ varies depending on the type of scalar field considered.
To leading order we can write \cite{louis}:
\be
\label{MatterMetric}
{\tilde K} = h(\tau_s) {\cal V}^{-a}
\ee
with the exponent $a\geq 0$ taking different values for the different
kinds of matter fields in the model. Here $h$ is a flavour dependent function
of the smaller moduli that in general will be very hard to compute.

We have found that the F-term contribution only cancels the leading
$m_{3/2}^2$, giving scalars suppressed by $\ln (m_{3/2})$, if $a=2/3$. 
This applies to D3 brane adjoint scalars and D7 Wilson lines. For
adjoint D7 matter, $a=0$ and the scalar masses are comparable to
$m_{3/2}$. Of course, Standard Model matter fields are in
bifundamental representations and should correspond to D3-D7 or D7-D7
matter. In this case the only calculations for $\mc{K}$ are in the
context of toroidal orbifolds, where $0 \le a < 2/3$. In this case there
is no cancellation in (\ref{scalarmass}) and the scalars are
comparable to the gravitino mass, with positive mass squared, and heavier than the gauginos by the
small hierarchy $\ln(m_{3/2})$.

This also allows us to say something about flavour universality.
In the exponentially large volume scenario, the physical picture is
that Standard Model matter is suppported on
almost-vanishing small cycles within a very large internal space ($\mc{V} \sim 10^{14}$).
The physics of flavour is essentially local physics which is determined by the
geometry of the small cycles and their intersections.
 Consequently, all flavours should see the
large bulk in the same way, as the distinctive flavour physics is local
not global. Therefore the power of $a$ in
(\ref{MatterMetric}) should be flavour-universal. The function
$h(\tau_s)$, in contrast, \emph{is} sensitive to the local geometry
and so should not be flavour-universal.

In these circumstances we can both show 
universality for the soft scalar masses and also
estimate the fractional level of non-universality. In the sum (\ref{scalarmass})
the leading $m_{3/2}^2$ term and the terms involving $F^b$ are
flavour-universal and give a universal contribution of
$\mc{O}(m_{3/2}^2)$. Universality fails due to the F-terms associated
with the small moduli. We can then rewrite (\ref{scalarmass}) as
\be
\label{Universality}
m_i^2 \sim \underbrace{\Bigg( m_{3/2}^2 + F^b \bar{F}^b \partial_b
\partial_{\bar{b}} \ln \tilde{K} \Bigg)}_{\hbox{universal}} + \underbrace{\left( \sum_s F^s \bar{F}^s
\partial_s \partial_{\bar{s}} \ln \tilde{K} \right)}_{\hbox{non-universal}}.
\ee
The $F^b F^{\bar{s}} + F^{\bar{b}} F^s$ cross-terms vanish.
As $F^s \sim \frac{m_{3/2}}{\ln(m_{3/2})}$ we obtain
\bea
\label{ScalarUniversality}
m_i^2 & \sim & \, m_{3/2}^2 (1+
 \epsilon_i), \nonumber \\
\Rightarrow m_i & \sim & m_{3/2} \left( 1 + \frac{\epsilon_i}{2} \right),
\eea
where non-universality is encoded in $\epsilon_i \sim \frac{1}{\ln (m_{3/2})^2}$.
As we require $m_{3/2} \sim 1 \hbox{TeV}$, we estimate the fractional
non-universality for soft masses as $\sim 1/(\ln (10^{18}/10^3))^2
\sim 1/1000$.

It is remarkable that these general
results can be extracted despite our ignorance of the precise
dependence of the K\"ahler potential on the matter fields. This is
possible because in the above scenario flavour physics is local while
supersymmetry breaking is global, and there exists a controlled
expansion in $\frac{1}{\mc{V}}$.

We also note that the large-volume scenario naturally addresses the
$\mu$ problem. This is because the natural scale for any mass term,
susy or non-susy, is $\mu \sim \frac{M_P}{\mc{V}} \sim 1
\hbox{TeV}$. Indeed the dilaton and complex structure moduli, which
are stabilised supersymmetrically by the fluxes, do acquire masses of
this order. Essentially this arises because the scalar potential has a
prefactor $e^{\mc{K}} \sim \frac{1}{\mc{V}^2}$, which sets the natural
scale for any mass term $\mu^2$.

The phenomenology of flux compactifications has been much studied
recently. 
The above discussions suggest new scenarios beyond those 
 previously considered in \cite{hepth0505076,aqs} and the KKLT case
  \cite{hepth0503216, hepph0504036, hepph0504037,
  hepph0507110, hepph0508029, hepph0509039, hepth0509158,
  hepth0511320, hepph0604192, hepph0604253}.
The most obvious case is that of an 
intermediate string scale and thus a TeV scale gravitino mass, 
with squarks and sleptons heavy and comparable
to the gravitino mass, while gauginos are suppressed by a $(\ln m_{3/2})$
factor. 
As the gaugino masses are suppressed, it is necesary to include 
anomaly mediated contributions in addition to the gravity-mediation
expressions above. However as the scalar masses are heavy and
comparable to the gravitino mass
the contribution of anomaly mediation is in that case negligible
- this is just as
well given the notorious problem of tachyonic sleptons for pure
anomaly mediation. It will be also interesting to analyse the
phenomenology of the non-universality predicted in equation
(\ref{ScalarUniversality}).
A detailed investigation of these and related scenarios is in progress \cite{ToCome}.

\section{Conclusions and Outlook}

This note has focused on the two complementary topics of gaugino and
scalar masses in the IIB string landscape. 
We first showed that suppressed gaugino masses
$M_a \sim \frac{m_{3/2}}{\ln(m_{3/2})}$
are a generic feature of the
landscape and occur whenever the stationary locus of a modulus $T$ is
purely determined by nonperturbative superpotential effects
$e^{-aT}$. This small hierarchy
has previously been identified in simple KKLT models.
Our results show that this extends to models with arbitrary numbers of
K\"ahler moduli and also to the large volume non-susy models of
\cite{hepth0502058, hepth0505076} in which the AdS minimum is already
non-supersymmetric and the gaugino mass hierarchy is the consequence
of a subtle cancellation.

For the large volume non-susy models we have verified this explicitly by
 performing an exact
calculation of the gaugino and moduli masses for compactifications on
$\mbb{P}^4_{[1,1,1,6,9]}$, obtaining the expected suppression as
 well as the numerical prefactor.

We also studied soft scalar masses in the exponentially large volume
scenario.
An important deviation from the KKLT scenario is that in this case
scalar masses are generically \emph{not} suppressed compared to the
gravitino mass. In this scenario there is thus a small hierarchy
between the scalar and gaugino masses,
\be
m_i \sim m_{3/2} \sim \ln (m_{3/2}) M_a.
\ee
For gaugino masses, anomaly mediation and gravity mediation are
therefore equally important, but for scalar masses gravity mediation
dominates and anomaly mediation is suppressed by the ordinary $1/(16
\pi^2)$ factor. This is interesting as it bypasses the standard problems of
tachyonic anomaly-mediated scalar masses.
A further difference from the KKLT scenario is that in the
large-volume models the relevant high-scale for an RG analysis is the
intermediate rather than GUT scale. This is because $m_{3/2} \sim
\frac{M_P W_0}{\mc{V}}$ while $m_s \sim \frac{M_P}{\sqrt{\mc{V}}}$,
and for $W_0 \sim 1$ weak-scale soft terms require $\mc{V} \sim
10^{14}$, giving an intermediate string scale. In this respect it will
be interesting to extend the results of \cite{aqs} to include this
small scalar-gaugino mass hierarchy.

Possibly the most interesting result of this paper 
is that the exponentially large volume
scenario naturally gives approximate flavour universality. This has been one of
the main problems for gravity mediated scenarios, but in our scenario
it comes
out naturally. Roughly, this arises
because the physics of flavour is local and hence insensitive to the
dominant F-term which is that associated to the overall volume.
Flavour can however see the suppressed F-terms associated with the
small cycles, and the same factor of
$\ln (m_{3/2})$ responsible for the F-term suppression also determines the
magnitude of non-universal contributions to soft masses.
As this scenario can also solve the hierarchy problem through the
dynamical volume stabilisation at exponentially large volumes, we find
it phenomenologically appealing. A more detailed further study of the
results of this paper will appear in \cite{ToCome}.

There is one final note of caution. In attempting to build realistic
models, there are sound reasons to
suppose that not all K\"ahler moduli are stabilised by
nonperturbative effects. In particular, if all moduli were stabilised
by nonperturbative effects then their axionic parts would all also be
heavy and a QCD axion capable of solving the strong CP problem would
not exist. If the K\"ahler modulus corresponding to the QCD cycle
is partially stabilised through perturbative effects, it is possible that
gluinos may be heavier than the remaining gauginos. 
It is therefore also interesting to analyse the phenomenology of a mixed
scenario with an ordering $m_{3/2} \sim m_i \gtrsim m_{\tilde{g}} >
m_{\tilde{W}} \sim \frac{m_{3/2}}{\ln(m_{3/2})}$.

\acknowledgments
We acknowledge useful conversations on the subject of this paper with 
S. Abdussalam, B. Allanach, C.P.  Burgess,  D. Cremades and K. Suruliz.
JC is funded by EPSRC and Trinity College, Cambridge. FQ is partially
funded by PPARC and a Royal Society Wolfson award.

\end{document}